\documentclass[twocolumn]{IEEEtran}
\usepackage{cite}
\usepackage{amsmath,amssymb,amsfonts}
\usepackage{hyperref}
\usepackage{graphicx}
\usepackage{textcomp}
\usepackage{multicol}
\usepackage{multirow}
\usepackage{subfigure}
\usepackage{xcolor}
\usepackage[normalem]{ulem}

\usepackage{algorithm}
\usepackage{algpseudocode}
\algnewcommand\algorithmicinput{\textbf{Input:}}
\algnewcommand\Input{\item[\algorithmicinput]}
\algnewcommand\algorithmicoutput{\textbf{Output:}}
\algnewcommand\Output{\item[\algorithmicoutput]}

\begin{document}
\title{Robust Detection of Surgical Tools in Endoscopic Videos}
\title{Robust Surgical Tools Detection in Endoscopic Videos with Noisy Data}
\author{Adnan Qayyum$^{1}$, Hassan Ali$^{1}$, Massimo Caputo$^2$, Hunaid Vohra$^2$, Taofeek Akinosho$^3$, Sofiat Abioye$^3$, \\ Ilhem Berrou$^3$, Paweł Capik$^3$, Junaid Qadir$^4$, and Muhammad Bilal$^{3,*}\thanks{Corresponding Author: muhammad.bilal@uwe.ac.uk}$ \\
$^1$Information Technology University, Lahore, Pakistan \\
$^2$Bristol Heart Institute, University of Bristol, Bristol, United Kingdom \\ 
$^3$University of the West of England, Bristol, United Kingdom \\
$^4$Qatar University, Doha, Qatar \\
}
% \thanks{Submitted: 04 July 2021
% This paragraph of the first footnote will contain the date on 
% which you submitted your paper for review. It will also contain support 
% information, including sponsor and financial support acknowledgment. For 
% example, ``This work was supported in part by the U.S. Department of 
% Commerce under Grant BS123456.'' 
% }
% \thanks{Adnan Qayyum, Waqas Sultani, Fahad Shamshad, and Junaid Qadir are with the Information Technology University of the Punjab, Lahore, Pakistan (e-mail: adnan.qayyum@itu.edu.pk, waqas.sultani@itu.edu.pk, fahad.shamshad3@gmail.com, and junaid.qadir@itu.edu.pk). }
% \thanks{Rashid Tufail is with Wajahat Surgical Hospital, Attock, Pakistan (e-mail: rashidtufail0@gmail.com).}
% }

\maketitle

\begin{abstract}
Over the past few years, surgical data science has attracted substantial interest from the machine learning (ML) community. Various studies have demonstrated the efficacy of emerging ML techniques in analysing surgical data, particularly recordings of procedures, for digitizing clinical and non-clinical functions like preoperative planning, context-aware decision-making, and operating skill assessment. However, this field is still in its infancy and lacks representative, well-annotated datasets for training robust models in intermediate ML tasks. Also, existing datasets suffer from inaccurate labels, hindering the development of reliable models. In this paper, we propose a systematic methodology for developing robust models for surgical tool detection using noisy data. Our methodology introduces two key innovations: (1) an intelligent active learning strategy for minimal dataset identification and label correction by human experts; and (2) an assembling strategy for a student-teacher model-based self-training framework to achieve the robust classification of 14 surgical tools in a semi-supervised fashion. Furthermore, we employ weighted data loaders to handle difficult class labels and address class imbalance issues. The proposed methodology achieves an average F1-score of 85.88\% for the ensemble model-based self-training with class weights, and 80.88\% without class weights for noisy labels. Also, our proposed method significantly outperforms existing approaches, which effectively demonstrates its effectiveness.

\end{abstract}

\begin{IEEEkeywords}
Surgical tool detection, \and surgical data science, \and ensemble learning, \and active learning
\end{IEEEkeywords}

\section{Introduction}
\label{sec:introduction}
% \IEEEPARstart{I}{n}

%In recent years, surgical data science has emerged as a promising discipline within surgical science, advocating the adoption of various machine learning (ML) techniques for intermediate tasks in surgical scene understanding. 

In recent years, surgical data science has emerged as a promising discipline within the field of surgical science, promoting the adoption of data-driven methods such as machine learning (ML) and deep learning (DL) techniques. These advanced approaches have been instrumental in enhancing surgical scene understanding and tackling a wide range of intermediate tasks in surgery, including object detection (tools, tasks, hands), surgical workflow analysis, and tissue segmentation for visual risk tracking \cite{maier2017surgical}. Surgical tool detection plays a crucial role, enabling downstream applications such as preoperative planning, surgical skill assessment, and task detection \cite{liu2022towards}. Moreover, it can be used to develop an automated surgical skills assessment system that can provide objective feedback on the dexterity of the practitioner's surgical procedures, which can be used for further improvement \cite{jin2018tool}. However, despite the significant attention given to surgical tool detection, only a few studies have thoroughly analyzed the robustness of DL approaches in the presence of issues such as class imbalance, and label noise.

% \textcolor{red}{\sout{recent years, surgical data science has emerged as a promising discipline in surgical science. In particular, different tasks have been modelled using various machine learning (ML) and deep learning (DL) techniques such as surgical tool detection, surgical scene understanding, semantic segmentation of surgical images, and surgical tool tracking. Surgical tool detection and tracking is an important task that enables various downstream tasks, e.g., surgical skills assessment and surgical task detection, etc. In the literature, various DL-based approaches have been presented for surgical tool detection, however, only a few of them considered analysing proposed methods through a robustness lens (for which they only considered class imbalance issue). }}

\begin{figure*}[!ht]
    \centering
    \includegraphics[width=\textwidth]{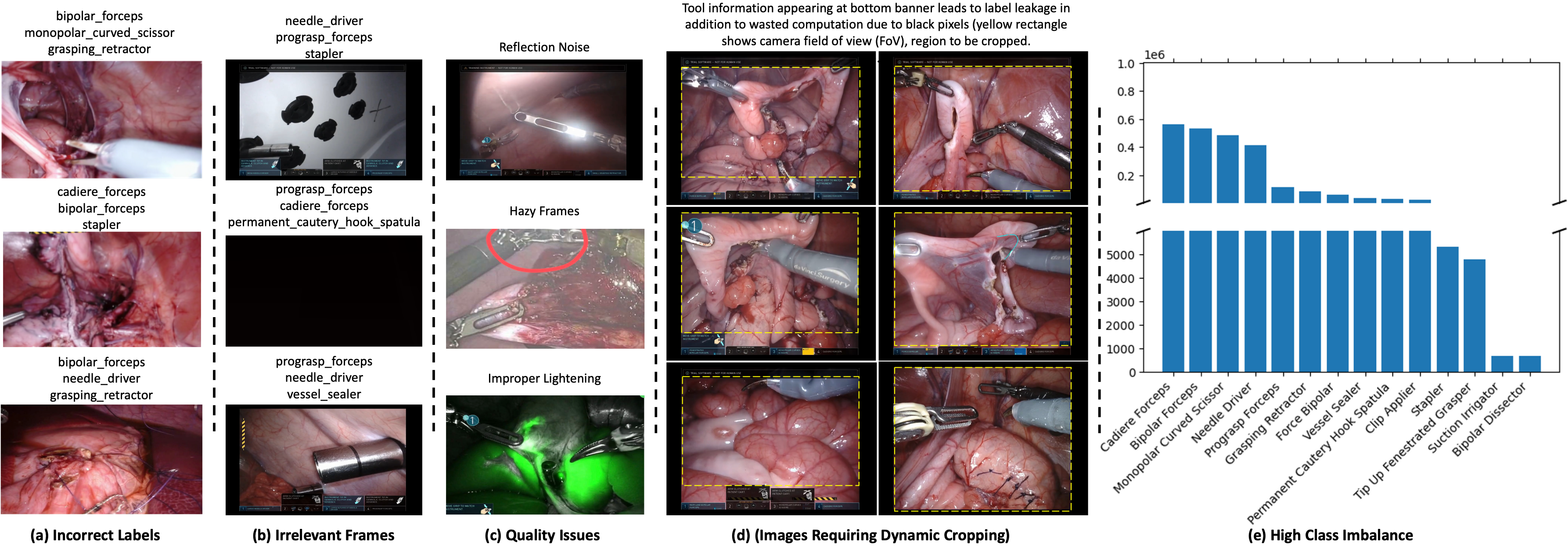} 
    \caption{An illustration of surgical data quality issues in our dataset including incorrect labels, irrelevant frames, image quality issues, images leaking label information, and high class imabalance. \textit{Our proposed methodology is capable of training a robust model in the presence of these issues.}}
    \label{fig:data_noises}
\end{figure*}

However, unlike other well-established disciplines, surgical data science is still in the developmental stage and lacks high-quality representative datasets necessary for developing robust applications to digitize clinical and non-clinical tasks \cite{qayyum2023can}. Surgical datasets, such as endoscopic videos, commonly suffer from various data quality issues, including device-related noise, improper lighting, label leakage, class imbalance, and label noise (as shown in Fig. \ref{fig:data_noises}). Currently, most available datasets are insufficient for efficient and large-scale model training and contain imperfections. The creation of curated and well-annotated benchmark datasets for surgical tool detection poses significant challenges. As annotating surgical videos is exceptionally demanding due to factors such as limited expert availability, time requirements, and wide diversity of surgical interventions and tools across different procedures.

% \textcolor{red}{\sout{However, unlike other disciplines, surgical data science is still developing and lacks good-quality representative datasets that could be used to develop and validate ML/DL-based approaches. Moreover, it is very common for endoscopic videos to be afflicted by various issues such as improper lighting, the presence of spurious correlations that could influence the underlying model to learn non-robust features, class imbalance, inherent sources of noise, etc. Also, currently, available datasets are not sufficient for the efficient and large-scale training of DL models and contain imperfections. On the other hand, the creation of a clean and gold-standard dataset for surgical tool detection is quite challenging. As the annotation of surgical videos (such as endoscopic videos) is quite challenging due to various factors such as expert availability, the time required, and the diversity of surgical interventions and tools used in different procedures.}}

The development of robust models is imperative for the successful translation of ML-powered predictive products in clinical care \cite{qayyum2020secure}. To achieve state-of-the-art (SOTA) performance, DL models rely on large amounts of clean data with high-quality annotations. However, annotation of medical data is quite challenging, costly, and time-consuming \cite{qayyum2022single}. Also, as previously mentioned, surgical datasets are often plagued by various data quality issues. In many cases, these datasets contain annotations with different types of noise, leading to incorrect descriptions of surgical scenes. Training robust models on such noisy datasets necessitates the use of a systematic methodology that incorporates diverse strategies for data preparation, model training, error analysis, and deployment. 

In this paper, we attempt to address the challenge of creating robust surgical tool detection models on challenging data characterized by high class imbalance and significant label noise. Note that we refer to the term robustness as the ability of the model to effectively learn relevant features from noisy data without significantly compromising its learning capabilities and predictive performance. Specifically, we utilize a dataset of approximately 24,694--30-second recordings of surgical robot procedures, where one label per video is provided to indicate the presence of surgical tools. However, the surgical scene can change significantly within 30-second interval, resulting in significant label noise when extrapolating labels over frames, as the presence of tools in each frame does not accurately represents the video-level labels. Manual label correction of such a dataset, consisting of approximately 4.2 million images is impractical. Moreover, we propose a novel ensembling strategy for developing student-teacher models, enabling the learning of the underlying mapping function ($f: x \rightarrow y$) from the noisy data, where $x$ and $y$ denote input and output, respectively. To the best of our knowledge, this is the first study to comprehensively tackle these challenges that employ active learning (AL) together with integrating ensembling in self-training to train a robust model using noisy data for surgical tool detection. Specifically, this paper makes the following salient contributions.

% \textcolor{red}{\sout{In general, surgical videos are annotated on the video level and annotations are then extrapolated to frames for training of DL models. This approach can potentially result in different kinds of noises such as label noise, label leakage, label misspecification, etc. In this paper, we attempt to model surgical tool detection using such data, which contains high-class imbalance and a significant amount of label noise. To the best of our knowledge, this is the first paper that uses student-teacher model-based self-training for learning underlying mapping function $f:$ $x \rightarrow y$ using noisy data. Where $x$ is the input and $y$ represents input and label by the models. Specifically, the following are the major contributions of this paper. }}

\begin{enumerate}
    \item We present a dataset containing more than 24k images that have been manually annotated for 14 different surgical tools. 
    \item We present an AL-based strategy to manually label surgical frames with minimal human effort.
    \item We present a student-teacher framework that is based on an ensemble model (containing four DL models) for surgical tool detection using noisy data. 
    \item We present the use of weighted data loaders (WDLs) for training student models in self-training framework to address high class imbalance issue. 
    \item We perform extensive analysis to validate the efficacy of the proposed framework and benchmark individual models as well as the ensemble model. 
\end{enumerate}

The rest of the paper is organized as follows. Section \ref{sec:related} presents a review of related articles. Section \ref{sec:method} describes the data and various preprocessing techniques applied to the data along with the proposed method for robust detection of surgical tools using noisy data. The results are described in Section \ref{sec:results} and the paper is concluded in Section \ref{sec:concs}.

\section{Related Work}
\label{sec:related}

Various approaches have been proposed for surgical tool detection, ranging from classical ML to DL-based solutions. Bouget et al. \cite{bouget2015detecting} introduced a two-stage method that leverages the local appearance of surgical tools at the pixel level and enforces global shape using tool-specific shape templates. Their findings emphasize the importance of intermediate semantic labelling for achieving robust detection performance. Kumar et al. \cite{kumar2013product} explored the use of classical image processing techniques, including point-based, region-based, and optical flow, for surgical tool detection and tracking. Richa et al. \cite{richa2011visual} proposed a weighted mutual information-based image similarity function for visual tracking of surgical tools, specifically for proximity detection in retinal surgeries. 

A number of existing research studies have used CNN for surgical tool detection. In \cite{jin2018tool}, authors proposed a region-guided CNN model for surgical tool detection and tracking that was used for surgical skills assessment by evaluating tools' movement, usage, range, and motion. Their method was the first attempt towards spatial localization of surgical tools in laparoscopic surgical videos. Similarly, Liu et al. \cite{liu2022towards} proposed a depth-wise separable convolution operation that was used to develop a convolutional LSTM model for surgical tool detection. The use of reinforcement learning to control positive and negative sample adaptation during the training model for surgical tool detection is presented in \cite{wang2020surgical}. Garc{\'\i}a-Peraza-Herrera et al. \cite{garcia2016real} formulated the surgical tool detection problem as a segmentation problem and tracked it using optical flow. In their work, a fully convolutional network was used as the segmentation network; however, their method can distinguish between different tools.

In another work  \cite{ciaparrone2020comparative}, Ciaparrone et al. employed the mask R-CNN model for the segmentation of surgical tools and evaluated 12 different backbone CNN architectures. Twinanda et al. \cite{twinanda2016endonet} proposed a CNN model for surgical tool presence detection. The authors extracted features from a fully connected layer of the trained CNN model and then used the support vector machine model and the Hierarchical hidden Markov model for surgical phase detection. The proposed method was validated using two endoscopic surgical databases, i.e., Cholec80 and EndoVis. Hasan et al. \cite{hasan2021detection} proposed a novel framework named augmented reality tool network (ART-Net) for the detection, segmentation, and 3D rendering of surgical tools in endoscopic videos. The proposed framework is an integration of CNN architecture (with one encoder and multiple decoders) and algebraic geometry, which are collectively used to perform the aforementioned tasks. In \cite{shi2020real}, Shi et al. proposed a CNN-based framework that incorporates coarse and refined detection modules. Furthermore, they integrated an attention module into the refined detection module that enforces the network to learn important features for surgical tool detection. Yang et al. \cite{yang2021efficient} proposed a CNN-based model for surgical tool detection that works by generating ghost feature maps by exploiting intrinsic feature maps. 

To address the problem of data imbalance, Jaafari et al. \cite{jaafari2021towards} employed different data augmentation techniques (such as rotation of different angles, mirroring, shearing, and padding) while preserving the tool's presence. The authors then fine-tuned a CNN (i.e., inception ResNet V2 pre-trained on ImageNet dataset) model using augmented data for the surgical tool classification task. To address the challenge of the availability of annotated data for surgical tool detection, Ali et al. \cite{ali2022semi} presented a student-teacher-based self-supervised learning framework that works by only utilising a small fraction of labelled data. Labelled data is used for training the teacher model, which is then inferred using unlabeled images to get pseudo labels for student model training. In addition, the authors integrated a region proposal network for the extraction of the region of interest in input images. A weakly supervised framework named pseudo-supervised surgical tool detection (PSTD) that incorporates three phases for pseudo-label generation is presented in \cite{xue2022new}. Specifically, to model the contextual information, PSTD employs a bi-directional adaptation weighting mechanism in the surgical tool detection classifier.

A novel modulated anchor network that works in conjunction with the Faster R-CNN model for surgical tool detection is presented in \cite{zang2019extremely}. The key purpose of the anchor network is to predict the spatial location of anchor shapes used in tools for training the backbone network. Furthermore, they proposed to incorporate a relation module in the network (one module after each fully connected layer) to model the relationship of a tool in a given image with other tools. Choi et al. \cite{choi2017surgical} present the utilisation of a SOTA object detection model, specifically You Only Look Once (YOLO), for surgical tool detection. However, only three surgical videos acquired at 25 fps were used for the evaluation of the model. We refer interested readers to comprehensive surveys that are focused on surgical tool detection for getting more detailed information about different DL-based methods \cite{wang2021visual,rivas2021review,bouget2017vision}.

Ensembling has been found to be significant for surgical tool detection, as highlighted in various studies. For example, in \cite{alshirbaji2021deep}, the authors proposed an ensemble model that uses VGG16 and ResNet50 models for spatial feature learning and two LSTM units on top of CNN models for temporal feature learning. Similarly, \cite{mishra2017learning} suggested a combination of CNN and stacked LSTM for spatial feature extraction and temporal information encoding. Jaafari et al. \cite{jaafari2022impact}, surgical tool detection was formulated as a multi-label classification problem. To address this, they utilised an ensemble model comprising three CNN architectures: Inception v-4, VGG-19, and NASNet-A. They used various data augmentation techniques to address the data imbalance problem and improve model training. The dataset in this study comprises only 80 videos acquired at a frame rate of 25 and has seven labels. Wang et al. \cite{wang2017deep} also formulated surgical tool detection as a multi-label classification problem and proposed an ensemble model that uses model averaging to ensemble predictions of trained GoogleNet and VGGNet. Our approach is similar to \cite{jaafari2022impact} and \cite{wang2017deep}, as they have also used ensembling. However, our study differs in the following ways: (1) our dataset comprises more samples; (2) the number of classes (i.e., tools) in our data is 14; (3) we consider learning from noisy data; and (4) we incorporated ensembling in self-training.

% Error analysis driven network modification for surgical tools detection in laparoscopic frames \cite{yin2022error}

% AGNet: attention-guided network for surgical tool presence detection \cite{hu2017agnet}

% Detection and correction of specular reflections for automatic surgical tool segmentation in thoracoscopic images \cite{saint2011detection} 

% Lapformer: surgical tool detection in laparoscopic surgical video using transformer architecture \cite{kondo2021lapformer}

% A contextual detector of surgical tools in laparoscopic videos using deep learning \cite{namazi2022contextual}

% Multi-label classification of surgical tools with convolutional neural networks \cite{prellberg2018multi}

% Graph convolutional nets for tool presence detection in surgical videos \cite{wang2019graph} 

% Oriented Localization of Surgical Tools by Location Encoding \cite{xue2021oriented} 

% Recognition and prediction of surgical actions based on online robotic tool detection \cite{park2021recognition}

% Analyzing laparoscopic cholecystectomy with deep learning: automatic detection of surgical tools and phases \cite{sahu2020analyzing}

% Surgical tool classification in laparoscopic videos using convolutional neural network \cite{alshirbaji2018surgical}

\section{Methodology}
\label{sec:method}
In this section, we present our proposed methodology for surgical tool detection using noisy data. Specifically, we start by first defining the problem and describing the dataset.  

\subsection{Problem Formulation}
% \textcolor{red}{\sout{We extract our dataset $\mathcal{D}$ from a collection of endoscopic surgical video (ESV) clips comprising of different minimally invasive surgeries performed by surgeons containing using various surgical tools. Each clip $c_j$ in the ESV collection is comprised of multiple frames $c_j = \bigcup_{i=0}^{f_j-1} \{x_j^{(i)}\}$ acquired at a frame rate of 60 frames per second (fps), where $f_j$ denotes the total number of frames in $c_j$. Further, each ESV clip has been assigned a set of three unique labels $y_j = \{y_{j1}, y_{j2}, y_{j3}\}$ by the dataset providers, where each of the three labels is an instance of the label set $L = \bigcup_{i=0}^{T-1} \{t_i\}$ where $T=14$ denotes the total number of tools (i.e., classes) present in the dataset. Therefore, $\mathcal{D} = \bigcup_{j=0}^{n-1} \{(c_j, y_j)\} = \bigcup_{j=0}^{n-1} \bigcup_{i=0}^{f_j-1} \{(x^{(i)}_j, y_j)\}$, where $n=24694$ is the total number of ESV clips. For simplicity, we denote our dataset as $\mathcal{D} = \bigcup_{i=0}^{N-1} \{(x_i, y_i)\}$, where is $x_i \in \mathcal{R}^{1280\times720\times3}$ is a single frame and $l_i \in L$ is the set of labels assigned to the ESV clip from which $x_i$ is sampled, where $N$ denotes the dataset size.}}

Our objective is to develop a robust model for surgical tool detection using endoscopic surgical video (ESV) images, ESV data contains data imperfections and noisy labels. ESV exhibit varying tool visibility within the field of view (FoV) throughout the duration of video clips. The extrapolation of annotations provided at the ESV level introduces inconsistencies between the ground truth labels and the actual tools present in the view, resulting in significant label noise. This label noise poses a unique challenge for deep neural networks (DNNs) in effectively classifying the tools present in the image. Our objective is to address this challenge by predicting the presence of tools in each frame $x_i$. We formulate the surgical tool detection task as a multi-label image classification problem, where the goal is to predict three out of the fourteen tools that are present in each frame.

We extract our dataset $\mathcal{D}$ from a collection of ESV clips comprising different minimally invasive surgeries performed by surgeons using various surgical tools. Each clip $c_j$ in the ESV collection consists of multiple frames $c_j = \bigcup_{i=0}^{f_j-1} \{x_j^{(i)}\}$ acquired at a frame rate of 60 frames per second (fps), where $f_j$ denotes the total number of frames in $c_j$. Furthermore, each ESV clip has been assigned a set of three unique labels $y_j = \{y_{j_1}, y_{j_2}, y_{j_3}\}$ by the dataset providers. Each of the three labels is an instance of the label set $L = \bigcup_{i=0}^{T-1} \{t_i\}$, where $T=14$ denotes the total number of tools (i.e., classes) present in the dataset. Therefore, our dataset can be represented as $\mathcal{D} = \bigcup_{j=0}^{n-1} \{(c_j, y_j)\} = \bigcup_{j=0}^{n-1} \bigcup_{i=0}^{f_j-1} \{(x^{(i)}_j, y_j)\}$, where $n=24,694$ is the total number of ESV clips. For simplicity, we denote our dataset as $\mathcal{D} = \bigcup_{i=0}^{N-1} \{(x_i, y_i)\}$, where is $x_i \in \mathcal{R}^{1280\times720\times3}$ represents a single frame and $l_i \in L$ represents the set of labels assigned to the ESV clip from which $x_i$ is sampled. Here, $N$ denotes the dataset size.

% \textcolor{red}{\sout{As tools visible in the field of view (FoV) of endoscopic images keep changing throughout the duration of the clips, static annotation of the clips (such as the one provided) results in several inconsistencies between the assigned ground truths and the true labels. This problem is more commonly known as the label noise, which makes it considerably more challenging for DNNs to classify the tools effectively. Our objective is to predict the tool(s) present in each frame $x_i$ of the endoscopic videos independently. We pose the surgical tool detection task as a multi-label image classification problem. Note that in each frame, the task is to predict only three (out of the fourteen) tools that are present at a time.}}

\subsection{Data Description and Preprocessing}
\subsubsection{Data Description}

% \textcolor{red}{\sout{The dataset used in this study has been taken from a recent Medical Image Computing and Computer Assisted Intervention (MICCAI) challenge named ``Surgical Tool Localization in Endoscopic Vidoes'' \cite{zia2023surgical}. The dataset comprises a total of 24,694 ESV clips, recorded at 60fps, which have been annotated via crowdsourcing. Annotators were asked to assign labels to the surgical clips based on the instrument displayed on the user interface appearing at the bottom of the videos. The dataset poses a noticeable challenge due to the presence of various kinds of noise, including incorrect tool labels, blank frames, and the user interface at the bottom of the videos. The primary reason for the incorrect labels is the extrapolation of labels of video clips across frames, which resulted in assigning wrong labels in many cases, significantly affecting the model's learning capability. This is because instruments specified for video clips were not present in the given frame. Additionally, most of the videos contain leaked instrument labels through the user interface controls shown at the bottom of the video, i.e., a tool being shown in the user interface was not in use during the surgical phase. The distribution of fourteen different tools in the dataset is shown in Fig. \ref{fig:tool_dist}. It can be seen that the dataset is highly imbalanced which makes it challenging to develop a robust model that can good performance in the presence of high-class imbalance and label noise. }}

The dataset used in this study is sourced from the ``Surgical Tool Localization in Endoscopic Videos'' challenge at the Medical Image Computing and Computer Assisted Intervention (MICCAI) 2022 conference \cite{zia2023surgical}. The dataset comprises a total of 24,694 endoscopic surgical video (ESV) clips captured at a frame rate of 60fps. The provided annotations are obtained programmatically from the robotic system, indicating the tools deployed or removed at specific timestamps. However, it is important to note that these labels do not guarantee an accurate representation of the tools present in the FoV. As a result, the dataset exhibits various data quality issues (see Fig. \ref{fig:data_noises}), including weak labels, blank frames, noisy labels, and label leakages. Label leakage occurs when the user interface of the robotic system at the bottom of ESV images reveals the names of the deployed tools, further complicating the modeling task. The challenge of inaccurate labels is exacerbated when weak labels at the ESV level are extrapolated across frames, leading to incorrect labels in many cases. This introduces significant difficulty for the models in learning robust features for surgical tool detection, as the instruments specified in the ESV clips may not be present in every video frame. Additionally, Fig. \ref{fig:data_noises} (e) presents the distribution of the fourteen surgical tools within the dataset, highlighting its highly imbalanced nature, which poses an additional challenge in training a robust model for surgical tool detection.

\subsubsection{Data Preprocessing} In our proposed method, we have performed intuitive data preprocessing to remove different types of noises and to ensure good-quality data and present a systematic methodology for learning from noisy data containing label noise. Next we discuss the various techniques that are used for the preprocessing of videos and labels. 

\textbf{a) Videos Preprocessing:} The following are the key steps involved in preprocessing of the surgical videos. 

\textit{Frames Sampling:} As noted previously, our key objective is to predict the tool(s) present in each frame of the ESV clip independently. To achieve this, we sample the frames of each ESV clip in the first preprocessing step. On average, an ESV clip contains 1800 frames, which makes a total of over 44 million frames, where each frame $x_i \in \mathcal{R}^{1280\times720\times3}$ exceeds the size of 1 TB on the hard drive. To overcome this issue, we sample the frames at the rate of 10Hz which provides a sufficient number of samples for model training and allows us to execute our experiments in a reasonable time. 

\textit{Dynamic Cropping using Segmentation:} We apply dynamic cropping over the sampled frames to extract the region of interest in the camera FoV. To achieve this, we first create a custom dataset by randomly sampling a frame from each of the ESV clips. We then use the Prodigy\footnote{https://prodi.gy/} tool and a custom algorithm to annotate these sampled frames and then train a U-Net-based dynamic segmentation model over the custom dataset to distinguish the foreground (region containing useful information) and the background regions (irrelevant pixels, i.e., user interface banner) in the frames. Our learned segmentation model efficiently removes the background region comprising the user interface control panel that appears at the bottom; the disclaimer notice at the top; and the black borders appearing on the left and right sides of the videos (see Fig. \ref{fig:data_noises}(d)). 

As these extra pixels can only contribute towards prolonged network training and wasted computation while also potentially negatively impacting the predictive performance of the trained models \cite{ilyas2019adversarial}. In addition, the presence of tool information (at bottom banner) leads to label leakage issue, which the model being trained can use as a shortcut \cite{geirhos2020shortcut}. More details about our dynamic segmentation model and results can be seen in our preliminary work on addressing label leakage in ESV \cite{qayyum2023segcrop}. 

\textit{Black Frames Removal:} In the dataset, a few videos contain entirely black frames without any visual information and a few videos contain partially black frames, while they have also received labels due to extrapolating their parent ones. We have removed such frames as they can potentially affect the model performance.

\textbf{b) Labels Preprocessing:}
We performed exploratory data analysis to find some prominent labelling noise in the dataset and fixed the identified issues using the following methods.  

\textit{Labels Standardisation:} We used lambda expressions to clean label strings, i.e., removing different symbols that include brackets ([]), quotes (''), hyphens (-), slashes (\textbackslash\textbackslash). Also, the underscore (\_) symbol was removed with white space.

\textit{Label Extrapolation:} The dataset provides one label for each video, while the expected algorithm is required to predict labels for instruments' presence for all the frames in the video. Therefore, video labels are required to be extrapolated for frames. This extrapolation introduced a lot of noise in the training labels. There are many examples where the label might indicate three instruments' presence but only two or fewer tools can be seen in the given frame. This is often the case when the surgeon moved the tool out of view even though it is installed on the robotic system. Learning from noisy labels is recognised as one of the unique challenges in this competition presented.

\begin{figure*}[!t]
    \centering
    \includegraphics[width=0.98\textwidth]{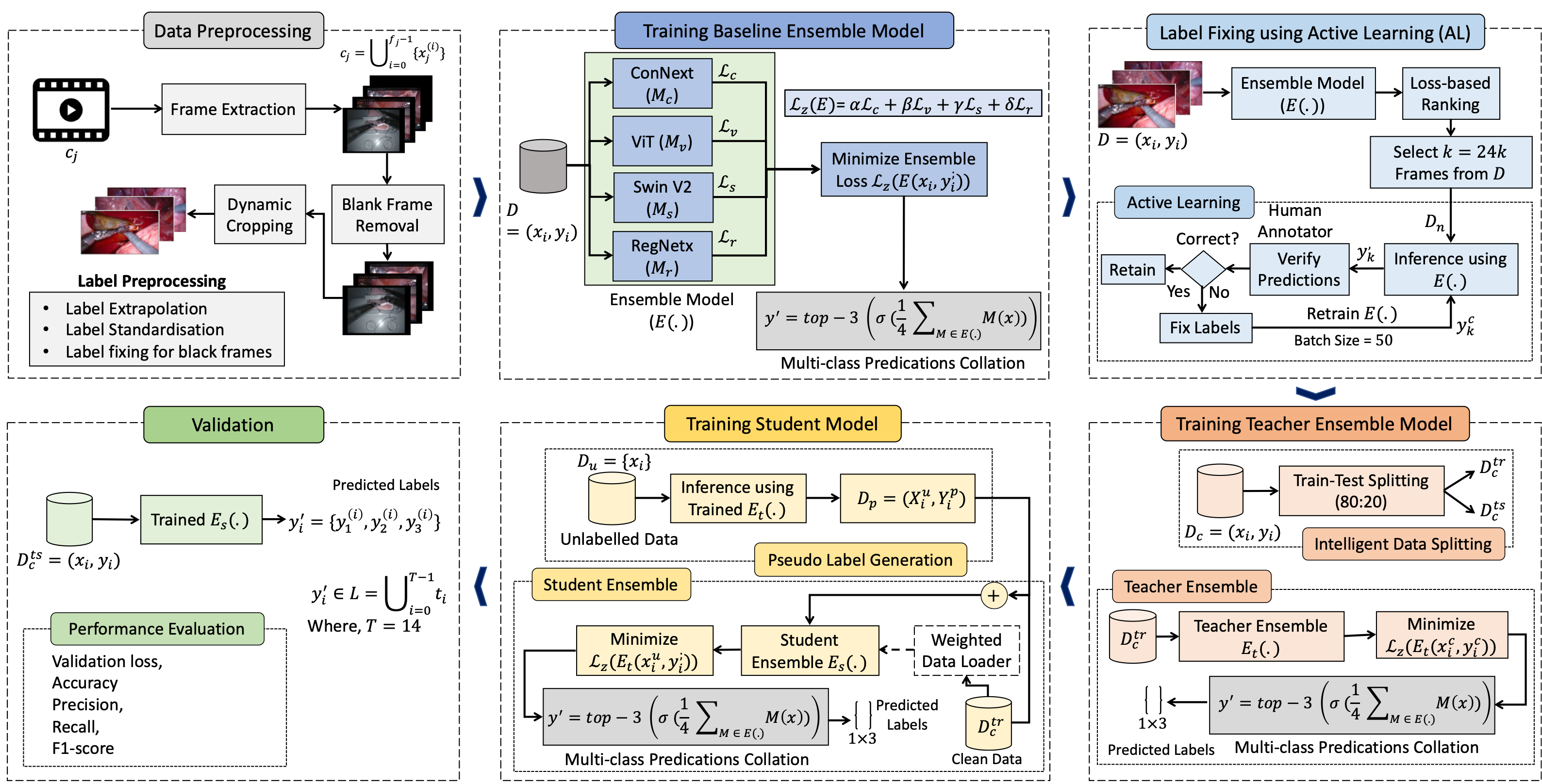} 
    \caption{Proposed methodology for robust detection of surgical tools with noisy data. \textit{Firstly,} we perform data preprocessing to resolve some prominent data imperfections. \textit{Secondly,} we train the baseline ensemble model, and then we use loss-based epistemic scoring for minimal dataset selection for label fixing. \textit{Thirdly,} we perform active learning (AL)-based manual label correction. \textit{Fourthly,} we train teacher ensemble model $E_t(\cdot)$ to get the pseudo labels. \textit{Fifthly}, these pseudo labels are used for training the student ensemble model $E_s(\cdot)$. \textit{Finally,} we evaluate the performance of the proposed student ensemble model.  }
    \label{fig:method}
\end{figure*}

\subsection{Training Baseline Model(s)}

To establish our baseline results for surgical tool detection, we trained our ensemble model while employing different data augmentation methods to develop an effective model (that can be used for minimal data identification for manual label cleaning). In ML practice, it is widely recognized that employing data augmentation techniques is crucial for efficient training of DL models. With this in mind, we explored various data augmentation approaches to improve our ensemble training. Specifically, we tested three techniques: squishing, cropping, and padding. Our findings indicated that squishing images yielded superior results compared to cropping and padding. Furthermore, we also investigated the impact of rectangular versus squared resizing. We discovered that our baseline models achieved better performance when the dataset was scaled to maintain a similar aspect ratio as the original dataset (having a rectangular dimensions of 1280$\times$720). This insight informed our modeling choice of preserving the aspect ratio during the process for optimal model performance. 

% In addition to applying data augmentation techniques, we employed progressive resizing in the training of our baseline models. With progressive resizing, we initially trained the models using small images that were 4 times smaller in size than the original images. The dimensions of the images were then progressively increased until reaching the maximum size (while maintaining aspect ratio), typically the size of the original images. For the final model, we trained it on the actual image dimensions for a reduced number of epochs. We performed 12 epochs for training on the small images and then used only 6 epochs for each of the two successive resized datasets. This approach allowed our baseline models to improve their accuracy by 1.35\%. However, it is important to note that training multiple models using progressive resizing comes at a higher computational cost, making it practically infeasible for ESV images, which can be quite large. The ensemble model design is described next.

% \textit{Ensemble Model Design:} 

\textit{Ensemble model $E(\cdot)$:} Our ensemble model $E(\cdot)$ comprises four different models $E(\cdot)=$ $\{M_c(\cdot),$ $ M_v(\cdot),$ $ M_s(\cdot),$ $ M_r(\cdot)\}$, respectively representing ConvNext \cite{liu2022convnet}, ViT \cite{dosovitskiy2020image}, Swin V2 \cite{liu2022swin}, and RegNext \cite{radosavovic2020designing} models. Each of these models is selected intuitively based on their superior performance in modelling similar tasks.\footnote{\href{https://www.kaggle.com/code/jhoward/the-best-vision-models-for-fine-tuning}{\scriptsize{https://www.kaggle.com/code/jhoward/the-best-vision-models-for-fine-tuning}}}

\textit{Ensemble loss $\mathcal{L}_z(\cdot)$}: We then use $\mathcal{D}$ to train $E(\cdot)$ using an ensemble loss function $\mathcal{L}_z(\cdot)$, defined below:

\begin{multline}
    \mathcal{L}_z ( E(x), y ) = \alpha \mathcal{L} ( M_c(x), y ) + \beta \mathcal{L} ( M_v(x), y )  \\
    + \gamma \mathcal{L} ( M_s(x), y ) + \delta \mathcal{L} ( M_r(x), y )
\end{multline}

where $x,y$ denote the input and the corresponding ground truth, respectively and $\mathcal{L}(\cdot)$ denotes the standard multi-class classification loss of the DNN over $\mathcal{D}$, respectively, and $\alpha, \beta, \gamma$ and $\delta$ are the regularization hyperparameters tuned based on the predictive confidence (loss) of each model.

Aforementioned models were selected using grid search optimization for architectural search to design ensemble learners. Specifically, we explore different neural network architectures from various families including ResNet, ConvNext, Swin V2, EfficientNets, ResNexts, REGNETX, and ViT. We selected the final model to be an ensemble of four pre-trained models (i.e., ConvNext, Swin V2, ViT, and REGNETX) from various families.
% based on their efficacy to predict the presence of tools in surgical videos. 
We performed exhaustive experimentation to see which model families can learn the given classification tasks with greater accuracy. We found that ConvNext and Swin outperform all other architectures. Besides, ViTs and REGNETX were also equally performing well. So we choose to use four different models across these four families. Also, for initial testing, we used smaller architectures to choose the right families and then tested their larger versions for ensemble learning. We found that larger architectures were not improving the overall performance but instead were massively overfitting on the given task. Therefore, we decided to use the smaller architectures instead.

\subsection{Proposed Ensemble Based Self-Training Method}

In this section, we will discuss our proposed self-training (student-teacher model learning)-based strategy for the robust detection of surgical tools using noisy labels. Our proposed ensemble learning approach for surgical tool detection using self-training is presented in Figure \ref{fig:method}. The following three key steps are involved in the proposed self-training framework: (1) creating clean (human-)labeled data using active learning;  (2) training student-teacher ensemble models using a self-supervised learning strategy; and (3) using WDLs to ensure fair learning and address high-class imbalance issues. We will discuss these steps in detail below.

% starting with the ensemble model used in both active learning and self-supervised learning paradigms.

\subsubsection{Label Cleaning using Active Learning}
Firstly, we perform an extensive exploratory data analysis to find the enormity of label noise in our dataset $\mathcal{D}$. Our analysis revealed that the original labels assigned at the video level also contain significant label noise, which was then amplified due to the extrapolation of labels from video to frame level. Therefore, to address the noisy labels issue, we first create a clean dataset $\mathcal{D}_c$ that is a small subset of the complete dataset $\mathcal{D}$, where $|\mathcal{D}_c| \ll |\mathcal{D}|$. Specifically, we proposed an ensemble model $E(\cdot)$ that is iteratively trained using AL paradigm for fixing labels to develop $\mathcal{D}_c$, which consists of three different steps.
% the details about $E(\cdot)$ are provided below. 
% To compile $\mathcal{D}_c$, we leverage an AL paradigm by iterating over three different steps.

\textit{In the first step}, we train a baseline ensemble model on $\mathcal{D}$ and then use the trained $E(\cdot)$ to automatically identify and manually fix potentially mislabeled input samples. 

\textit{In the second step}, we identify a set of $k=24997$ samples from $\mathcal{D}$ on which the ensemble model $E(\cdot)$ exhibits the highest classification error (signifying that $E(\cdot)$ finds it hard to confidently predict these frames due to the presence of label noise), where $k \ll N$. Formally,

\begin{equation}
    \text{Repeat $k$ times}: \mathcal{D}_c = \mathcal{D}_c \cup \max_{x_i \in \mathcal{D}, x_i \notin \mathcal{D}_c} \mathcal{L}_z( E(x_i), y_i )
\end{equation}

We then compute the output of $E(\cdot)$, which is a set of the maximum probability assigned to each class $t_i \in L$ by any model $M \in E$.

\begin{equation}
    E(x) = \bigcup_{t_i \in L} \max_{M \in E} \{ M (t_i | x) \}
\end{equation}

where $L$ denotes the label set (as defined previously).

\textit{In the third step}, we then ask two human experts to verify the model predictions $E(x)$ for all $x \in \mathcal{D}_c$. If the model predictions are found to be incorrect, the experts manually suggest correct predictions and $\mathcal{D}_c$ is accordingly updated. Note that initially, manual annotations were done by our clinical partners and then a team of two annotators was trained by clinical experts to perform the rest of the annotations. All annotations were then validated by clinical experts to eliminate or fix any discrepancies. The updated $\mathcal{D}_c$ is then used to (re)train $E(\cdot)$, note that a batch size of $50$ was used for fine-tuning $E(\cdot)$. The above process is repeated several times to let $E(\cdot)$ continue to improve its performance on hard samples as the noisy labels get fixed, until the size of $\mathcal{D}_c$ becomes 24k. 

Finally, the compiled clean data $\mathcal{D}_c$ is divided into two distinct non-overlapping sets: the clean training data $\mathcal{D}_c^{tr}$ (comprising 80\% of $\mathcal{D}_c$) and the clean test data $\mathcal{D}_c^v$ (comprising the remaining 20\% of $\mathcal{D}_c$).

% \hassan{There must be some basis for labeling the ESV clips the way they have been labeled. It is likely that the ESV clips have been annotated based only on the observations of the initial few frames. We can also use these initial few frames as a starting point of $\mathcal{D}_c$.} \aq{Actually, we did an extensive analysis on the label noise, and we will add all details into the paper.}

\subsubsection{Student-Teacher Formulation for Ensemble Learning}
We proposed to leverage student-teacher formulation (i.e., self-supervised learning also known as self-training) to automatically mitigate the effect of noisy labels in the surgical tool detection task. Our proposed ensemble learning approach for surgical tool detection using self-training (i.e., student-teacher model) works in two steps, as presented in Fig. \ref{fig:method}. 

\textit{In the first step}, we train teacher ensemble model $E_t(\cdot)$ having same architecture as $E(\cdot)$ defined previously. We train $E_t(\cdot)$ over $\mathcal{D}_c^{tr}$ using $\mathcal{L}_z(\cdot)$, and evaluate $E_t(\cdot)$ over $\mathcal{D}_c^v$. Given an input $x$, the output of $E_t(\cdot)$ is calculated as sigmoid of mean non-probabilistic outputs (logits) of all models $M \in E_t(\cdot)$,

\begin{equation}
    E_t(x) = \sigma \left( \frac{1}{4} \sum_{M \in E_t(\cdot)} M(x) \right)
\end{equation}

\textit{In the second step}, we re-label the unclean data $(\mathcal{D} - \mathcal{D}_c)$ by first querying $E_t(\cdot)$ with $(\mathcal{D} - \mathcal{D}_c)$ to get the pseudo labels $E_t(x), \forall x \in (\mathcal{D} - \mathcal{D}_c)$, and use these pseudo labels to create a pseudo labelled dataset $\mathcal{D}_s = \{(x_i, E_t(x_i))\}, \forall x \in (\mathcal{D} - \mathcal{D}_c)$. 

Finally, we define a student ensemble model $E_s(\cdot)$ (having same architecture as $E(\cdot)$), and train $E_s(\cdot)$ over the augmented dataset $\mathcal{D}_s \cup \mathcal{D}^{tr}_c$. As before, the output of $E_s(\cdot)$ at the inference time is computed by passing the average non-probabilistic outputs of all models $M \in E_s(\cdot)$ through the sigmoid function and using the top-3 labels as the final classification decision,

\begin{equation}
    E_s(x) = \operatorname{top-3} \left( \sigma \left( \frac{1}{4} \sum_{M \in E_s(\cdot)} M(x) \right) \right)
\end{equation}

\begin{algorithm}[t]
    % \footnotesize
    \scriptsize
    \caption{Methodology}
    \label{alg:methodology}
    \begin{algorithmic}[1]
    
    \Input
    \Statex $\mathcal{D} = \{(x_i, l_i)\}_{i=0}^{N-1} \gets$ training data of size $N$
    \Statex $M_c, M_v, M_s, M_r \gets$ individual models
    \Statex $\mathcal{L}_z(\cdot) \gets$ ensemble loss function
    
    \Output
    \Statex $E_s = \{M_c, M_v, M_s, M_r\} \gets$ trained ensemble model

    \Statex
    \Procedure{Ensemble Training}{$E_{in}, \mathcal{D}_{in}$}
        \State \verb|/... ensemble training code .../|
    \EndProcedure

    \Statex\Statex \textcolor{blue}{//Active learning-based label cleaning}
    \State $\mathcal{D}_c \gets \{\}$ :empty set
    \While{$|\mathcal{D}_c| \leq 24000$}
        \State $E \gets \text{copy } \left( \{M_c, M_v, M_s, M_r\} \right)$
        \State $E \gets$ Ensemble Training ($E, (\mathcal{D}-\mathcal{D}_c) \cup \mathcal{D}_c$)
        \State $\mathcal{L} \gets \{\}$ :empty set
        \For{$(x,y) \in (\mathcal{D}-\mathcal{D}_c)$}
            \State $\mathcal{L} \gets \mathcal{L} \cup \mathcal{L}_z(E(x), y)$
        \EndFor
        \For{i = [0..500] }
            \State $(x_m, y_m) \gets \text{Correct} \max_{x_i \in \mathcal{D}, x_i \notin \mathcal{D}_c} \mathcal{L}_z( E(x_i), l_i )$
            \State $\mathcal{D}_c \gets \mathcal{D}_c \cup (x_m, y_m)$
        \EndFor
    \EndWhile
    \State $(\mathcal{D}_c^{tr}, \mathcal{D}_c^v) \gets$ Split $(\mathcal{D}_c)$
    
    \Statex\Statex \textcolor{blue}{//Teacher model training using clean dataset $\mathcal{D}_c$}
    \State $E_t \gets \text{copy } \left( \{M_c, M_v, M_s, M_r\} \right)$
    \State $E_t \gets$ Ensemble Training $(E_t, \mathcal{D}_c^{tr})$
    
    \Statex\Statex \textcolor{blue}{//Student model training using pseudo dataset $\mathcal{D}_s$}
    \State $\mathcal{D}_s \gets \{\}$ :empty set
    \For{$(x,y) \in (\mathcal{D}-\mathcal{D}_c)$}
        \State $\mathcal{D}_s \gets \left( x, E_t(x) \right)$
    \EndFor
    \State $E_s \gets \{M_c, M_v, M_s, M_r\}$
    \State $E_s \gets$ Ensemble Training $(E_s, \mathcal{D}_s)$

    \Statex\Statex\textcolor{blue}{//Performance evaluation}
    \State Validate $E_s$ on $\mathcal{D}_c^v$
    
    \Statex
    \State \textbf{return} $E_s$
    \end{algorithmic}
\end{algorithm}

\subsubsection{Augmenting Self-Training-based Ensemble Learning}
In addition to the noisy label issues, our dataset also suffers from significant class imbalance issues (as evident from Fig. \ref{fig:data_noises} (e)). To overcome this issue we proposed to use a WDLs strategy to train the ML models while reducing the data bias due to class imbalance. We fine-tuned the ensemble models for several epochs using a WDLs to improve performance. Also, to further augment the capacity of our self-training-based ensemble model in learning hard labels, we employed label smoothing, which is a widely used regularization technique to improve model generalizability and prevent overfitting. Label smoothing defines a soft distribution over classes instead of using hard targets, i.e., it allows models to relax label boundaries slightly. Therefore, instead of 1s and 0s, the loss function is engineered to use a number less than 1 for 1s and a number a bit more than 0 for all 0s in the encoding vector. In this way, label smoothing assigns probability $p$ to a correct class and uniformly distributes the remaining ($1-p$) to the rest of the classes. This enables the model to be more robust to the perturbations in the input while sacrificing a bit of predictive confidence. In the literature, label smoothing has been shown quite successful in improving the performance and generalizability of DL models using noisy datasets \cite{szegedy2016rethinking}.

\section{Results and Discussions}
\label{sec:results}

\subsection{Experimental Setup}

\textit{Intelligent Data Splitting:} We realized that randomly splitting frames into the training and testing set will allow frames of one video to end up in both sets which will also lead to data leakage. Therefore, we devised a customized logic to intelligently split the dataset into training and testing sets, where the frames from videos are allocated to either training or testing sets unless the labelled tool combination has only one video in the dataset. This ensures that each tool combination is present in both training and testing sets (to ensure fair validation of the models). Note that in our datasets, there are eight tool combinations with just one video. Using the aforementioned strategy, our dataset is divided into training and testing sets using a split of 80\% and 20\%, respectively. 

\textit{Hyperparamters Selection:} 
% \textcolor{red}{\sout{An exhaustive hyperparameter tuning was performed to choose the appropriate data augmentation types, learning rates, and architecture types. We applied squish-based data augmentations and the best learning rate was selected using the learning rate scheduler, which minimized the model's loss by scaling the magnitude of weight updates. As very low learning can result in very large training time and slow learning and a considerably large learning rate can lead to undesirable divergent behaviour. Therefore, to overcome this issue, we used a cyclic learning rate for model training (as proposed in \cite{smith2018disciplined}), which utilizes stochastic gradient descent with warm restarts to create a learning  schedule that combines an aggressive annealing schedule with periodic restarts. The learning rate of 1e-2 seemed to have worked very well for all types of models, where each model was trained for a maximum of 12 epochs with a batch size of 64. To ensure a fair comparison, all models were trained using the same choice of hyperparameters.}}
An exhaustive hyperparameter tuning process was conducted in training models using the W\&B library, taking into account data augmentation types, batch sizes, learning rates, and architecture choices. Through experimentation, it was determined that image squishing provided better results compared to cropping. Furthermore, the cyclic learning rate approach, as introduced by Smith et al. \cite{smith2018disciplined}, was adopted for selecting the best learning rate for training ensemble models. This approach combines stochastic gradient descent with warm restarts, allowing for an annealing schedule with periodic restarts. Among the different learning rates tested, a value of 1e-2 consistently yielded the best performance across the ensemble models. To facilitate efficient batch processing, a batch size of 64 was utilized, with n\_workers set to 8 for parallel processing. Additionally, mixed precision training (FP16) was employed, resulting in a 20\% improvement in computation time. The training process involved initially training the DNN models' heads using the fit-one-cycle method for 12 epochs. Subsequently, the entire model was unfrozen, and training was continued for an additional 12 epochs using smaller learning rates, such as slice (1e-2/400, 1e-2/4). This progressive training strategy allowed the models to refine their performance by focusing on fine-tuning the lower layers while maintaining the learned features in the higher layers. To ensure a fair comparison, all models were trained using the same choice of hyperparameters.

\subsection{Baseline Results}
% \textcolor{red}{\sout{To get the baseline results, we first divided our original (noisy) data $\mathcal{D}$ into training $D^tr$ and testing $D^ts$ sets using a split of 80\% and 20\%, respectively. We then trained our baseline models (which include Swin V2, ViT, ConvNext, and RegNetX) and ensemble model for the task of surgical tool detection using $D^tr$ in a fully supervised learning fashion. The trained model was then evaluated using test data, i.e., $D^ts$. Baseline results in terms of different performance metrics including validation loss, accuracy, precision, recall, and F1-score are summarized in Table \ref{tab:baseline}. The table highlights that our proposed ensemble model provided superior performance (in terms of accuracy (99.9\%) and precision (98.4\%)) as compared to four individual baseline models. Where, it can be seen that the performance of the ConvNext model is relatively less as compared to all other models (Swin V2, ViT, and RegNetX), while these three models provided a similar performance in terms of most of the metrics. Note that due to the presence of a significant amount of label noise so we can see that the model is outperforming on all metrics but visual inspection (performed manually) revealed that the predictions are not correct in most cases.}}

To establish a baseline for our experiments, we initially partitioned our original dataset, $\mathcal{D}$, into the training set $D^tr$ and testing set $D^ts$, utilizing an 80\% - 20\% split, respectively. Subsequently, we conducted fully supervised training of our baseline models, including Swin V2, ViT, ConvNext, and RegNetX, as well as an ensemble model, for the task of surgical tool detection using $D^tr$. The trained models were then evaluated on the test data, $D^ts$. Table \ref{tab:baseline} provides an overview of the baseline results, encompassing various performance metrics such as validation loss, accuracy, precision, recall, and F1-score. Notably, our proposed ensemble model outperformed the individual models in terms of accuracy (99.9\%) and precision (98.4\%). It is worth mentioning that the ConvNext model exhibited relatively lower performance compared to the other models (Swin V2, ViT, and RegNetX), while these three models showcased similar performance across most metrics. However, it is crucial to note that due to the presence of significant label noise, the models demonstrated strong performance in terms of the evaluation metrics, whereas manual visual inspection revealed that the predictions were often incorrect in many cases.

% Please add the following required packages to your document preamble:
% \usepackage{multirow}
% \begin{table}[!t]
% \centering
% \caption{Baseline results using original (noisy) data for surgical tool detection. }
% \label{tab:baseline}
% \scalebox{0.79}{
% \begin{tabular}{|c|l|c|c|c|c|c|}
% \hline
% \textbf{Tr. Size} & \multicolumn{1}{c|}{\textbf{Model}} & \textbf{Validation Loss} & \textbf{mAP} & \textbf{mAR} & \textbf{mAA} & \textbf{mAF1} \\ \hline
% \multirow{5}{*}{246,626} & ConvNext & 0.02471 & 0.96628 & 0.93587 & 0.99526 & 0.94968 \\ \cline{2-7} 
%  & RegnetX & 0.02833 & 0.97359 & 0.92547 & 0.99425 & 0.94676 \\ \cline{2-7} 
%  & VIT & 0.01998 & 0.97978 & 0.95983 & 0.99610 & 0.96920 \\ \cline{2-7} 
%  & Swin & 0.02173 & 0.97649 & \textbf{0.96780} & 0.99651 & \textbf{0.97196} \\ \cline{2-7} 
%  & Ensemble & \textbf{0.01369} & \textbf{0.98404} & 0.95724 & \textbf{0.99853} & 0.969401 \\ \hline
% \end{tabular}}
% \end{table}

\begin{table}[!t]
\centering
\caption{Baseline results using original (noisy) data for surgical tool detection. }
\label{tab:baseline}
\scalebox{0.75}{
\begin{tabular}{|l|c|c|c|c|c|}
\hline
\multicolumn{1}{|c|}{\textbf{Model}} & \textbf{Loss} & \textbf{mAP} & \textbf{mAR} & \textbf{mAA} & \textbf{mAF1} \\ \hline
 ConvNext & 0.02471 & 0.96628 & 0.93587 & 0.99526 & 0.94968 \\ \hline
 RegnetX & 0.02833 & 0.97359 & 0.92547 & 0.99425 & 0.94676 \\ \hline
 VIT & 0.01998 & 0.97978 & 0.95983 & 0.99610 & 0.96920 \\ \hline 
 Swin & 0.02173 & 0.97649 & \textbf{0.96780} & 0.99651 & \textbf{0.97196} \\ \hline
 Ensemble & \textbf{0.01369} & \textbf{0.98404} & 0.95724 & \textbf{0.99853} & 0.969401 \\ \hline
\end{tabular}}
\end{table}

\subsection{Results for Label Cleaning using AL}
% \textcolor{red}{\sout{Our active learning-based semi-automated annotation strategy resulted in the correction of noisy labels. As discussed above, we selected a bath size of 50, where our ensemble active learner was inferred to get the predictions. Whereas, incorrect labels were manually fixed by a human expert and then the model was retrained (fine-tuned) on the newly corrected labels. Figure \ref{fig:label_noise} presents an analysis in terms of the percentage of corrected (modified) labels that contain noise and were, therefore, influencing the model to give misclassifications. In addition, we also present an analysis based on the number of incorrect labels (i.e., predictions) that need to be corrected through manual intervention in Figure \ref{fig:reduction}. It is evident from the figure that our ensemble model was able to capture newly injected knowledge in the active learning paradigm, thereby, reducing the manual efforts required progressively, while the model learns to predict correctly.}}

\begin{figure}[!ht]
    \centering
    \includegraphics[width=0.3\textwidth]{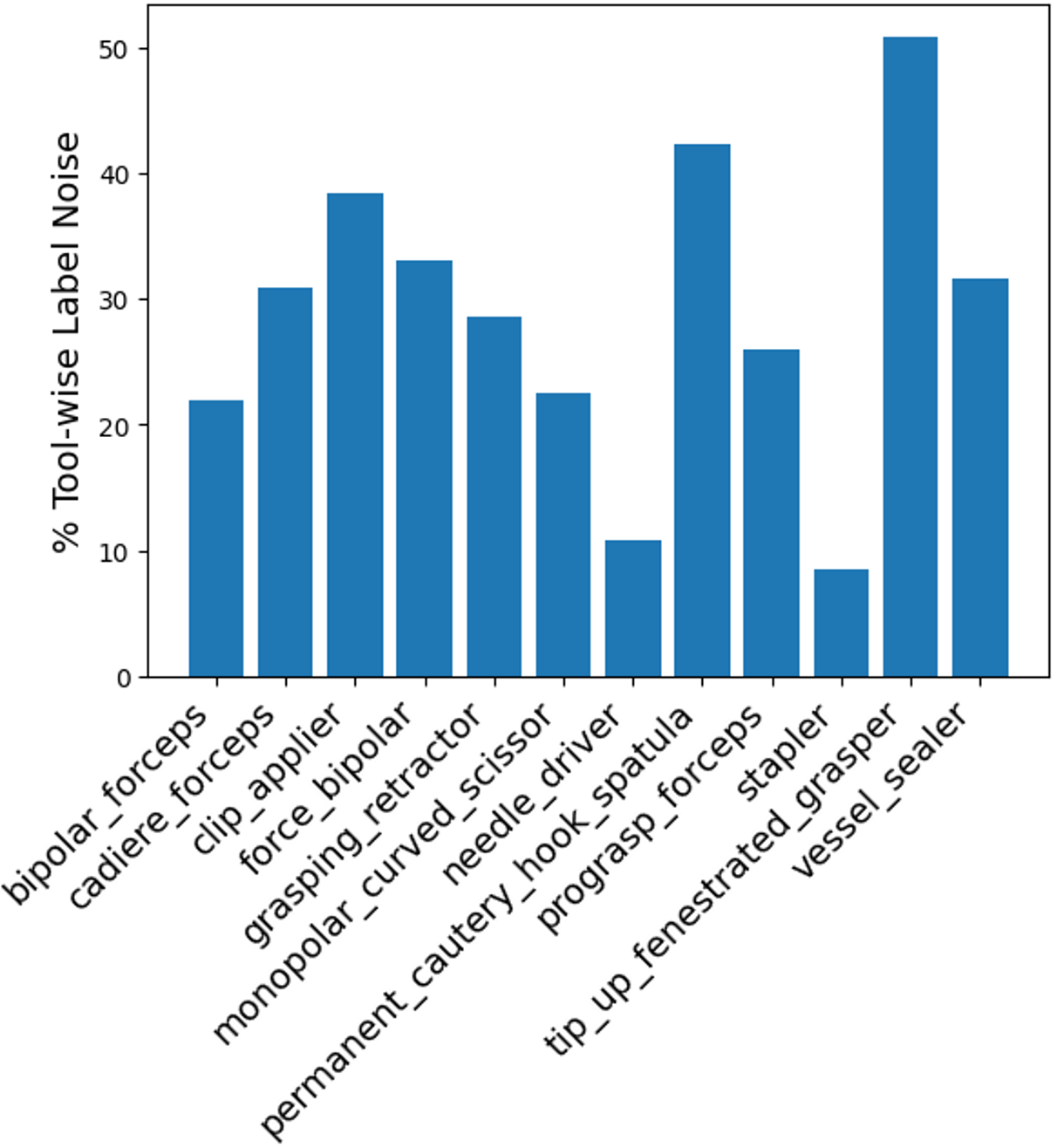} 
    \caption{Tool-wise percentage of corrected labels due to noise correction.}
    %Percentage correction of noisy labels (where the new bars represent the modified/corrected labels that were identified as misclassification).}
    \label{fig:label_noise}
\end{figure}

\begin{figure}[!ht]
    \centering
    \includegraphics[width=0.3\textwidth]{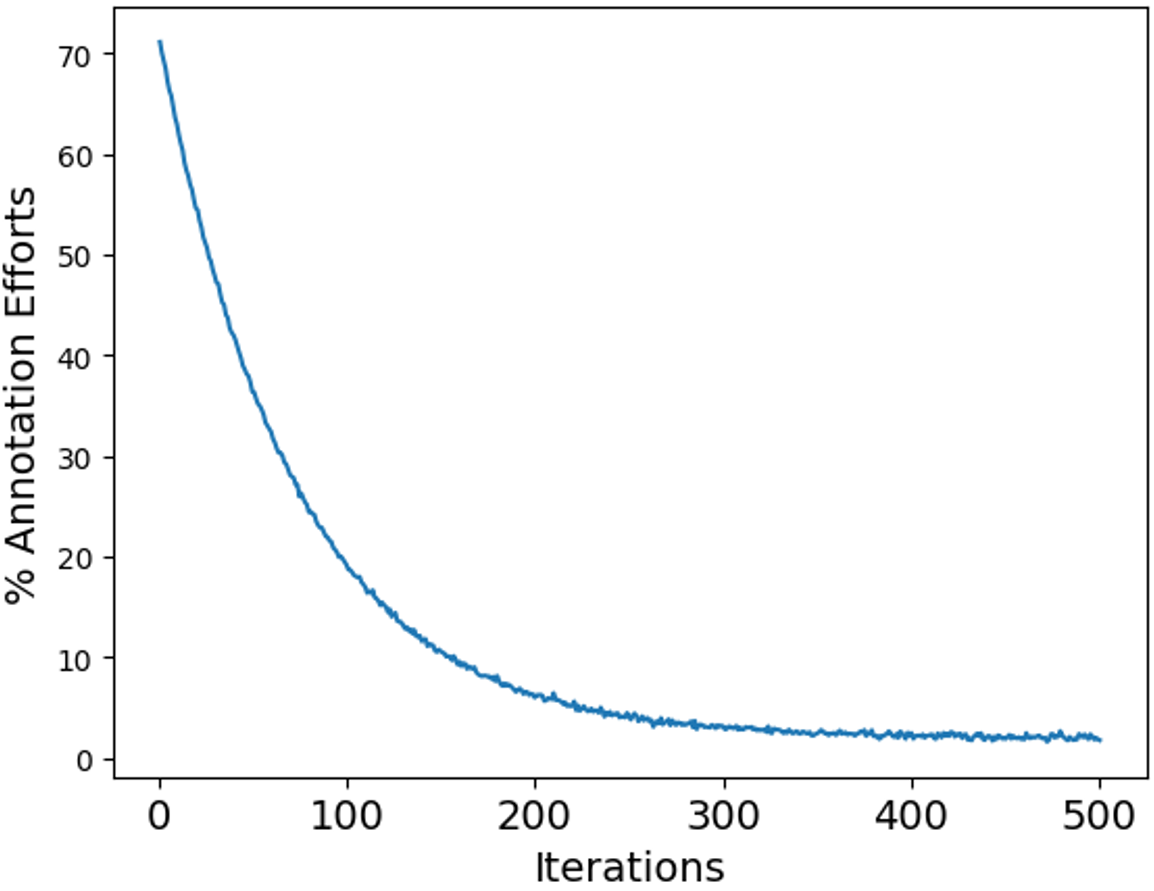} 
    \caption{Our weakly-supervised AL-based annotation strategy reduces the number of manual corrections (effort) over iterations.}
    %Analysis depicting improvement of active learning model in terms of the number of incorrect predictions over different iterations. Our active learning-based semi-automated annotation strategy progressively reduced the manual effort to just a few samples.}
    \label{fig:reduction}
\end{figure}

To address the challenge posed by significant label noise in our dataset, we employed a label-cleansing strategy. Given the impracticality of manually reviewing and rectifying labels for the entire dataset due to its large size, we employed a loss-based epistemic scoring method to identify a minimal subset (5\%) of the dataset for manual labelling. Our manual labelling process was facilitated by an AL-based semi-automated annotation strategy. We leveraged the Prodigy tool to implement our AL strategy. In the AL framework, human experts initially manually corrected the tool labels for the first batch of samples. This triggered the retraining of the baseline ensemble learner using the corrected samples, resulting in relatively accurate default labels for subsequent batches. Each time a batch was processed, the retraining process was automatically initiated, progressively improving the model's performance. Fig. \ref{fig:label_noise} provides insights into the proportion of noise present for each surgical tool that required correction during the labelling process. Additionally, we present an analysis based on the number of label corrections required through manual intervention in Fig. \ref{fig:reduction}. It is evident that as our label cleansing strategy advanced and we continued fine-tuning our ensemble model, the number of manual corrections required were significantly decreased, reducing the workload for human experts through AL.

% Results for AL-based semi-supervised annotation strategy are presented in Fig. \ref{fig:AL_results}, where 

% \begin{figure*}[!t]
%     \centering
%     \subfigure[]{\includegraphics[width=0.3\textwidth]{Figures/noise_correction_AL.png}\label{fig:label_noise}}
%     \subfigure[]{\includegraphics[width=0.3\textwidth]{Figures/efforts_correction_AL.png}\label{fig:reduction}}
%     \caption{Demonstrating the efficacy of AL-based manual data annotation with minimal human effort. Fig. \ref{fig:label_noise} illustrates the percentage of tool-wise labels noise, while Fig. \ref{fig:reduction} highlights the effectiveness of our weakly-supervised AL-based annotation strategy in reducing the number of manual corrections (effort) over iterations.}    \label{fig:AL_results}
% \end{figure*}

% Please add the following required packages to your document preamble:
% \usepackage{multirow}
\begin{table*}[]
\centering
\caption{Performance evaluation of four baseline models and ensemble models in self-supervised learning fashion (i.e., student-teacher model formulation) in terms of different performance metrics. }
\label{tab:st-results}
\scalebox{0.8}{
\begin{tabular}{|l||c|c|l|c|c|c|c|c|c|}
\hline
\multicolumn{1}{|c||}{\textbf{\begin{tabular}[c]{@{}c@{}}Model\\ Type\end{tabular}}} & \textbf{\begin{tabular}[c]{@{}c@{}}Labelling\\ Strategy\end{tabular}} & \textbf{Tr. Size} & \multicolumn{1}{c|}{\textbf{Model}} & \textbf{Train Loss} & \textbf{Loss} & \textbf{mAP} & \textbf{mAR} & \textbf{mAA} & \textbf{mAF1} \\ \hline
\multirow{5}{*}{\textbf{Teacher}} & \multirow{5}{*}{\begin{tabular}[c]{@{}c@{}}Clean \\ Labels\end{tabular}} & \multirow{5}{*}{24,997} & ConvNext & 0.01562 & 0.02539 & 0.76554 & 0.74104 & 0.99174 & 0.75198 \\ \cline{4-10} 
 &  &  & RegnetX & 0.01911 & 0.03018 & 0.75315 & 0.72520 & 0.98984 & 0.73597 \\ \cline{4-10} 
 &  &  & VIT & 0.01552 & 0.02304 & 0.77396 & 0.75509 & 0.99244 & 0.76366 \\ \cline{4-10} 
 &  &  & Swin & 0.01400 & \textbf{0.02273} & 0.76759 & 0.75541 & \textbf{0.99294} & 0.75917 \\ \cline{4-10} 
 &  &  & Ensemble & 0.01606 & 0.02533 & \textbf{0.80506} & \textbf{0.78418} & 0.99174 & \textbf{0.79269} \\ \hline \hline
\multirow{5}{*}{\textbf{Student}} & \multirow{5}{*}{Pseudo Labels} & \multirow{5}{*}{221,629} & ConvNext & 0.01413 & 0.01932 & 0.79554 & 0.76104 & 0.99174 & 0.76198 \\ \cline{4-10} 
 &  &  & RegnetX & 0.01711 & 0.02295 & 0.77315 & 0.74520 & 0.98984 & 0.74597 \\ \cline{4-10} 
 &  &  & VIT & 0.01352 & 0.01640 & 0.79396 & 0.77509 & 0.99244 & 0.77366 \\ \cline{4-10} 
 &  &  & Swin & 0.01100 & \textbf{0.01426} & 0.79759 & 0.78541 & \textbf{0.99294} & 0.76917 \\ \cline{4-10} 
 &  &  & Ensemble & 0.01394 & 0.01823 & \textbf{0.83457} & \textbf{0.82899} & 0.99174 & \textbf{0.80880} \\ \hline
\end{tabular}}
\end{table*}

\subsection{Self-Training-Based Ensemble Model Results}

In this section, we present the results of our proposed ensemble model-based self-training approach, which involves training two sets of models: a teacher model and a student model. The teacher model $\mathcal{M}_t$ was trained using clean data obtained using the AL strategy. The purpose of the $\mathcal{M}_t$ was to infer new (pseudo) labels from the model and discard original noisy tool labels. This pseudo-labelled dataset is used to train $\mathcal{M}_s$ and validate using the clean AL labels data. The teacher model $\mathcal{M}_t$ builds upon the models trained in the AL, while the student model $\mathcal{M}_s$ further extends $\mathcal{M}_t$ through transfer learning. The results of the student-teacher model-based surgical tool detection in terms of various performance metrics, are summarized in Table \ref{tab:st-results}. The table presents the performance of the proposed ensemble model as well as the individual models trained at this stage. It is evident that the ensemble strategy outperforms all individual models in both teacher and student settings. Notably, the results in Table \ref{tab:st-results} indicate a decline in performance metrics other than accuracy compared to our baseline results (reported in Table \ref{tab:baseline}). This highlights the importance of considering metrics beyond accuracy alone in evaluating surgical tool detection models. Additionally, the table demonstrates that the proposed student ensemble model, in the self-training setting performs comparatively better than the teacher model. This improvement is due to the generation of high-quality pseudo labels by $\mathcal{M}_t$ (while eliminating label noise), despite the significant difference in the number of samples used for training teacher and student models.

\subsubsection{Augmenting Self-Training using Label Smoothing}
% \textcolor{red}{\sout{Label smoothing is a well-known regularization technique, which is used to improve the generalisability of the models and to prevent them from getting overfitted. As described above, to augment the capacity of our proposed self-training-based ensemble learning approach, we incorporated label smoothing during training. The results of label smoothing for standalone models (i.e., training) and the ensemble model, when trained using student-teacher-based self-training, are presented in Table \ref{tab:label-smooth}. Note that label smoothing is only applied in the training of student models. Overall label smoothing provided promising results in handling noisy labels, however, we observed a little decrease in performance metrics. We anticipate that this performance drop was expected that mainly occurred due to the relaxing of decision boundaries, as hard labels were replaced with soft labels (the literature also documents this trade-off \cite{szegedy2016rethinking}). Moreover, from Table \ref{tab:label-smooth}, we can see that on average the performance of the ensemble model using label smoothing was slightly less as compared to other models, however, the ensemble model outperformed all other models in terms of F1-score, which is an important metric. Swin v2 outperformed all other models in terms of accuracy, precision, and recall, while ViT resulted in minimum validation loss.}}    

We investigated the effectiveness of various regularization techniques, including label smoothing, to enhance the generalizability of our proposed ensemble model. To implement label smoothing, we utilised Fastai’s LabelSmoothingCrossEntropy loss function during training. The results of applying label smoothing to both standalone models and the ensemble model trained using the student-teacher-based self-training approach are summarized in Table \ref{tab:label-smooth}. It is important to note that label smoothing was only applied during the training of the student models. Overall, label smoothing showed promising results in handling noisy labels, although there was a slight decrease in performance metrics. This performance drop can be attributed to the relaxation of decision boundaries when hard labels are replaced with soft labels, as documented in the literature \cite{szegedy2016rethinking}. From Table \ref{tab:label-smooth}, we observe that, on average, the ensemble model with label smoothing performed slightly worse compared to the other models. However, the ensemble model outperformed all other models in terms of the F1-score, which is a crucial metric. Swin v2 exhibited superior performance in terms of accuracy, precision, and recall, while ViT achieved the lowest validation loss.

\begin{table*}[]
\centering
\caption{Performance of comparison of using different strategies to improve model performance in self-training. The reported results are based on the student model trained using pseudo labels. }
\label{tab:label-smooth}
\scalebox{0.8}{
\begin{tabular}{|c||l|l|l|l|l|l|l|l|}
\hline
\textbf{Experiment} & \multicolumn{1}{c|}{\textbf{\begin{tabular}[c]{@{}c@{}}Labelling \\ Strategy\end{tabular}}} & \multicolumn{1}{c|}{\textbf{Tr. Size}} & \multicolumn{1}{c|}{\textbf{Model}} & \multicolumn{1}{c|}{\textbf{Loss}} & \multicolumn{1}{c|}{\textbf{mAP}} & \multicolumn{1}{c|}{\textbf{mAR}} & \multicolumn{1}{c|}{\textbf{mAA}} & \multicolumn{1}{c|}{\textbf{mAF1}} \\ \hline
\multirow{5}{*}{\textbf{\begin{tabular}[c]{@{}c@{}}Label \\ Smoothing\end{tabular}}} & \multirow{5}{*}{Pseudo labels} & \multirow{5}{*}{221,629} & ConvNext & 0.02467 & 0.79554 & 0.76104 & 0.99174 & 0.76198 \\ \cline{4-9} 
 &  &  & RegnetX & 0.02587 & 0.77315 & 0.74520 & 0.98984 & 0.74597 \\ \cline{4-9} 
 &  &  & VIT & \textbf{0.01958} & 0.79396 & 0.77509 & 0.99244 & 0.77366 \\ \cline{4-9} 
 &  &  & Swin & 0.02268 & \textbf{0.79759} & \textbf{0.78541} & \textbf{0.99294} & 0.76917 \\ \cline{4-9} 
 &  &  & Ensemble & 0.02320 & 0.78920 & 0.77857 & 0.98565 & \textbf{0.78741} \\ \hline \hline
\multirow{5}{*}{\textbf{\begin{tabular}[c]{@{}c@{}}Weighted \\ Data Loaders\end{tabular}}} & \multirow{5}{*}{Pseudo labels} & \multirow{5}{*}{221,629} & ConvNext & 0.01232 & 0.84578 & 0.83104 & 0.99568 & 0.79835 \\ \cline{4-9} 
 &  &  & RegnetX & 0.01395 & 0.82290 & 0.81520 & 0.99112 & 0.76922 \\ \cline{4-9} 
 &  &  & VIT & 0.01140 & 0.83645 & 0.82509 & \textbf{0.99823} & 0.78124 \\ \cline{4-9} 
 &  &  & Swin & \textbf{0.00926} & 0.84595 & 0.83541 & 0.99676 & 0.79978 \\ \cline{4-9} 
 &  &  & Ensemble & 0.01145 & \textbf{0.86569} & \textbf{0.84457} & 0.99725 & \textbf{0.85880} \\ \hline
\end{tabular}}
\end{table*}

\begin{table*}[]
\centering
\caption{Comparison with existing SOTA methods for surgical tool detection (Legend: FSL--Fully Supervised Learning, SSL--Self-Supervised Learning, AL--Active Learning, Self-Training (ST) and NR--Not Reported).}
\label{tab:comp}
\scalebox{0.74}{
\begin{tabular}{|l|p{6cm}|c|ccc|c|c|c|c|c|}
\hline
\multicolumn{1}{|c|}{\multirow{2}{*}{\textbf{Reference}}} & \multicolumn{1}{c|}{\multirow{2}{*}{\textbf{Method}}} & \multirow{2}{*}{\textbf{Type}} & \multicolumn{3}{c|}{\textbf{Data Description}} & \multirow{2}{*}{\textbf{Noise Type}} & \multirow{2}{*}{\textbf{mAP}} & \multirow{2}{*}{\textbf{mAR}} & \multirow{2}{*}{\textbf{mAF1}} & \multirow{2}{*}{\textbf{mAA}} \\ \cline{4-6}
\multicolumn{1}{|c|}{} & \multicolumn{1}{c|}{} &  & \multicolumn{1}{c|}{\textbf{Data}} & \multicolumn{1}{c|}{\textbf{\#Tools}} & \textbf{Sample Size} &  &  &  &  &  \\ \hline
Sahu et al. \cite{sahu2016tool}  & Used CNN for features extraction and random forest classifier for classification. & FSL & \multicolumn{1}{c|}{M2CAI 2016} & \multicolumn{1}{c|}{7} & NR & NR & 54.5 & NR & NR & NR \\ \hline
Raju et al. \cite{raju2016m2cai} & Presented an ensemble learning approach that uses GoogleNet and VGGNet for tool classification. & FSL & \multicolumn{1}{c|}{M2CAI 2016} & \multicolumn{1}{c|}{7} & NR & NR & 63.7 & NR & NR & NR \\ \hline
Jin et al. \cite{jin2018tool} & Proposed a region-guided faster R-CNN with VGG-16 as the base model for tools detection and tracking. & FSL & \multicolumn{1}{c|}{\begin{tabular}[c]{@{}c@{}}Extension of M2CAI \\ (with locations)\end{tabular}} & \multicolumn{1}{c|}{7} & Labelled: 2532 & NR & 81.8 & NR & NR & NR \\ \hline
Zhang et al. \cite{zhang2020surgical} & Proposed modulated anchoring network based on Faster R-CNN for tool detection and localization. & FSL & \multicolumn{1}{c|}{\multirow{2}{*}{\begin{tabular}[c]{@{}c@{}}Used dataset \\ provided by \\ Jin et al. \cite{jin2018tool}\end{tabular}}} & \multicolumn{1}{c|}{7} & Labelled: 2532 & NR & 69.6 & NR & NR & NR \\ \cline{1-3} \cline{5-11} 
Ali et al. \cite{ali2022semi} & Proposed a student-teacher-based ST with a loss function for surgical tool detection and localization. & SSL & \multicolumn{1}{c|}{} & \multicolumn{1}{c|}{7} & \begin{tabular}[c]{@{}c@{}}Labelled: 1\%, 2\%, \\ 5\%, and 10\%\end{tabular} & \begin{tabular}[c]{@{}c@{}}Class \\ Imbalance\end{tabular} & 46.89 & NR & NR & NR \\ \hline
\multirow{2}{*}{This Paper} & Proposed using AL with human input to create small clean label data and presented a student-teacher-based ST framework for training ensemble model(s). & \multirow{2}{*}{\begin{tabular}[c]{@{}c@{}}AL \\ \&\\  SSL\end{tabular}} & \multicolumn{1}{c|}{\multirow{2}{*}{\begin{tabular}[c]{@{}c@{}}SurgToolLoc \\ 2022 Challenge\end{tabular}}} & \multicolumn{1}{c|}{\multirow{2}{*}{14}} & \multirow{2}{*}{\begin{tabular}[c]{@{}c@{}}Labelled: 20,000 \\ Unlabelled: 2,46,626\end{tabular}} & \multirow{2}{*}{\begin{tabular}[c]{@{}c@{}}Label Noise \\ and\\ Class Imbalance\end{tabular}} & 83.36 & 82.89 & 80.88 & 99.17 \\ \cline{2-2} \cline{8-11} 
 & Incorporated class weights while training models in the proposed ensemble model-based SSL. &  & \multicolumn{1}{c|}{} & \multicolumn{1}{c|}{} &  &  & \textbf{86.57} & \textbf{84.46} & \textbf{85.88} & \textbf{99.73} \\ \hline
\end{tabular}}
\end{table*}

\subsubsection{Augmenting Self-Training using Weighted Data Loaders (WDLs)}
% \textcolor{red}{\sout{In addition to noisy labels, our dataset also suffers from a high-class imbalance that could result in the development of biased models that provides unfair predictions. To overcome this issue, we proposed using weighted data loaders that work by assigning class weights to each class that are calculated based on the class-wise sample population. We then trained all individual models and as well as ensemble models using class weights to address the class imbalance in our proposed student-teacher formulation-based self-training framework. The results of this analysis are summarized in Table \ref{tab:label-smooth}, it is evident from the table that weighted data loaders provided performance improvement of approximately 5\% in terms of average F1-score, 3\% in terms of precision, and 2\% in terms of recall for ensemble model. Moreover, the performance of all individual models was also improved by approximately 4-5\% in terms of different performance metrics that include precision, recall, and F1-score. Note that the weighted data loader technique is only applied in the training of student models. This indicates the efficacy of incorporating class weights during the training of student models in our proposed self-training framework that improved the fairness of the model(s).}}

In addition to dealing with noisy labels, our dataset is plagued by a significant class imbalance, which can lead to the development of a biased model (favouring the learning of features from dominant classes while disregarding minority labels). To address this issue, we introduced a WDLs strategy that assigns weights to labels based on their distribution in the dataset. This approach ensures that minority classes are given higher priority during batch selection by the data loaders during the training process. WDLs strategy is not used for training models from scratch but rather for fine-tuning existing models for a few additional epochs to specifically learn the minority classes. We applied this strategy to train both individual models and ensemble models within our student-teacher formulation-based self-training framework. The results of this analysis are summarised in Table \ref{tab:label-smooth}, which demonstrates the performance improvement provided by this method. On average, the ensemble model achieved a 5\% improvement in F1-score, a 3\% increase in precision, and a 2\% increase in recall when utilising WDLs. Similarly, the performance of all individual models was also improved by 4-5\% across various performance metrics. This underscore the effectiveness of incorporating weights during student models training within our self-training framework, resulting in improved model robustness and performance. Note that WDLs method is exclusively applied during the training of the student models.

\subsection{Comparison with Existing SOTA}

%In this section, we provide a comparison of our proposed ensemble model-based self-training framework with existing similar approaches. 
%We note that existing datasets containing surgical endoscopic videos are scarce, e.g., two famous benchmark datasets are the Cholec80 dataset which contains 80 surgical videos and the m2cai-2016-tool dataset which contains 15 laparoscopic surgical videos. 

% \textcolor{red}{\sout{It is worth noting that there is a scarcity of existing datasets containing surgical endoscopic videos. For example, two of the most well-known benchmark datasets are the Cholec80 dataset, which comprises 80 surgical videos, and the M2CAI-2016-tool dataset, which contains only 15 laparoscopic surgical videos. In this paper, we used the latest dataset that was provided by MICCAI 2022 challenge on surgical tool detection in endoscopic surgical videos, which contains $24,694$ videos and annotations for 14 tool, unlike previous datasets that only contains annotations for 7 tools. A quantitative comparison of our proposed approach with existing SOTA methods is presented in Table \ref{tab:comp}, note that authors reported results are used for the comparison. It is evident from the table that our proposed approach is efficient in terms of different performance metrics for a similar task with SOTA methods (despite the fact that our method considered learning from data containing significant label noise and class imbalance).}}

It is worth noting that there is a scarcity of existing datasets containing ESV data. For instance, two of the most well-known benchmark datasets, the Cholec80 dataset, and the M2CAI-2016-tool dataset consist of only 80 surgical videos and 15 laparoscopic surgical videos, respectively. In this paper, we utilized the latest dataset provided by the MICCAI 2022 challenge on surgical tool detection and localisation in ESV. This dataset includes a total of 24,694 videos and annotations for 14 tools, unlike previous datasets that only had annotations for 7 tools. We conducted a quantitative comparison of our proposed approach with existing SOTA methods, as presented in Table \ref{tab:comp}. It is important to note that the reported results from the respective authors were used for the comparison. From the table, it is evident that our proposed approach achieves high efficiency in terms of various performance metrics, despite the challenges of learning from data containing significant label noise and class imbalance.

% Our paper can be compared to the following papers. 

% \begin{enumerate}
%     \item Paper \cite{jin2018tool}: Major Contribution: Tools detection and tracking using a Region Guided VGG-16 model. They used  M2CAI 2016 Tool Presence Detection Challenge dataset that have labels for 7 tools. They release a new version of the dataset with labels in the form of spatial boxes for tools detection.  
%     \item Paper \cite{liu2022towards}: We used the ATLAS Dione dataset [17], which included
% 99 action video clips featuring ten surgeons performing six
% operational tasks on the da Vinci Surgical System (dVSS).
%     \item A semi-supervised Teacher-Student framework for surgical tool detection and localization \cite{ali2022semi} used MICCAI2016 data 
%     \item Surgical tools detection based on modulated anchoring network in laparoscopic videos \cite{zhang2020surgical} Used The Cholec80 dataset contains 80 surgical videos and miccai2016 dataset. 
% \end{enumerate}

% \section{Limitations and Future Work}
% \label{sec:limit}

\section{Conclusions}
\label{sec:concs}
% \textcolor{red}{\sout{In this paper, we consider the problem of learning from noisy data that contains a significant amount of label noise for the task of surgical tool detection in endoscopic surgical videos. Our proposed method has two major phases that include creating a clean dataset by manually labelling using active learning with minimal effort and using student-teacher formulation-based self-supervised learning for training surgical tool detection models (where we train an ensemble model that consists of four deep learning models). Specifically, the teacher model is trained using human-labelled clean data and the student model is trained using pseudo-labelled data, where pseudo labels are generated by the teacher model by considering unclean (noisy) data as unlabelled. Moreover, to address the challenge of high-class imbalance, we used class weights during the training of our models in self-supervised learning that provided a performance improvement of approximately 3-5\% in terms of different performance metrics. Our future work includes developing an unsupervised learning-based methodology for addressing the problem of learning from noisy data.}} 

In this paper, we addressed the problem of training robust models for surgical tool detection in endoscopic surgical videos using imbalanced datasets with significant label noise. Our proposed systematic methodology consisted of two main phases: creating a clean dataset from a minimal sample through active learning (AL) and training surgical tool detection models using a student-teacher formulation-based self-supervised learning approach. Ensembling techniques were employed throughout the methodology, incorporating baseline, AL, teacher, and student models. The teacher model was trained using a manually labeled subset of cleaned data, while the student model utilised a pseudo-labeled dataset generated by the teacher model, considering unclean (noisy) data as unlabeled examples. This ensembling approach proved effective in training robust models despite the presence of noisy labels in the dataset. To address the challenge of high-class imbalance, we introduced class weights during the self-supervised learning training process, resulting in a notable performance improvement of approximately 3-5\% across different performance metrics. Moving forward, our future work involves the development of an unsupervised learning-based methodology to address the challenge of learning from noisy data. By exploring unsupervised learning techniques, we aim to further enhance the robustness and performance of our models in the presence of diverse data quality issues.

\section*{Acknowledgements}
The authors would like to acknowledge support from UWE Bristol Vice-Chancellor’s Challenge Fund (IVA HEART).

\bibliographystyle{IEEEtran}

\end{document}